\documentclass[10pt, conference, letterpaper]{IEEEtran}
\IEEEoverridecommandlockouts
\usepackage{cite}
\usepackage{subfigure}
\usepackage{amsmath,amssymb,amsfonts}
\usepackage{algorithm}
\usepackage{algorithmic}
\usepackage{graphicx}
\usepackage{textcomp}
\usepackage{xcolor}
\usepackage{hyperref}
\usepackage{balance}
\usepackage{bm}
\usepackage{xcolor}
\usepackage{makecell, caption, adjustbox} % 添加到导言区
\usepackage{booktabs}
\usepackage{amssymb}
\usepackage{amsthm}
\usepackage{pifont}
\usepackage{multirow}
\usepackage{tabularx}
\usepackage{siunitx}        % 在导言区加入
\newcommand{\cmark}{\ding{51}}  
\newcommand{\xmark}{\ding{55}}

\newcommand{\eg}{\textit{e.g\@.}}
\newcommand{\ie}{\textit{i.e\@.}}

\newtheorem{definition}{Definition}
\newtheorem{theorem}{Theorem}
\hypersetup{
  colorlinks   = true,    % Colours links instead of ugly boxes
  urlcolor     = blue,    % Colour for external hyperlinks
  linkcolor    = blue,    % Colour of internal links
  citecolor    = blue      % Colour of citations
}

\newenvironment{icompact}{
  \begin{list}{$\bullet$}{
    \itemindent -.05em
    \parsep 0pt plus 1pt
    \partopsep 0pt plus 1pt
    \topsep 2pt plus 2pt minus 2pt
    \itemsep 0pt plus 1.3pt
    \parskip 0pt plus 2pt
    \leftmargin 0.13in}
      }
{\normalsize
\end{list}
}

\newcommand{\para}[1]{\vspace{2pt}\noindent{\textbf{#1}}\hspace{10pt}\vspace{0.1pt}}
\newenvironment{sloppypar*}
{\sloppy\ignorespaces}
{\par}
\def\BibTeX{{\rm B\kern-.05em{\sc i\kern-.025em b}\kern-.08em
    T\kern-.1667em\lower.7ex\hbox{E}\kern-.125emX}}

\begin{document}\sloppy

\title{\texttt{PrivTune}: Efficient and Privacy-Preserving Fine-Tuning of Large Language Models via Device-Cloud Collaboration}

\author{\IEEEauthorblockN{Yi Liu\textsuperscript{1}, Weixiang Han\textsuperscript{2}, Chengjun Cai\textsuperscript{2,*}, Xingliang Yuan\textsuperscript{3}, and Cong Wang\textsuperscript{1,*}}
	\IEEEauthorblockA{\textsuperscript{1}City University of Hong Kong; \textsuperscript{2}City University of Hong Kong (Dongguan); \textsuperscript{3}University of Melbourne \\ yiliu247-c@my.cityu.edu.hk, \{weixiang.han, chengjun.cai\}@cityu-dg.edu.cn,\\ xingliang.yuan@unimelb.edu.au, congwang@cityu.edu.hk.}
    
\thanks{\textsuperscript{*}Cong Wang and Chengjun Cai are co-corresponding authors.}}

\maketitle

\begin{abstract}
With the rise of large language models, service providers offer language models as a service, enabling users to fine-tune customized models via uploaded private datasets. However, this raises concerns about sensitive data leakage. Prior methods, relying on differential privacy within device-cloud collaboration frameworks, struggle to balance privacy and utility, exposing users to inference attacks or degrading fine-tuning performance. To address this, we propose \texttt{PrivTune}, an efficient and privacy-preserving fine-tuning framework via Split Learning (SL). The key idea of \texttt{PrivTune} is to inject crafted noise into token representations from the SL bottom model, making each token resemble the $n$-hop indirect neighbors. \texttt{PrivTune} formulates this as an optimization problem to compute the optimal noise vector, aligning with defense-utility goals. On this basis, it then adjusts the parameters (\ie, mean) of the $d_\chi$-Privacy noise distribution to align with the optimization direction and scales the noise according to token importance to minimize distortion. Experiments on five datasets (covering both classification and generation tasks) against three embedding inversion and three attribute inference attacks show that, using RoBERTa on the Stanford Sentiment Treebank dataset, \texttt{PrivTune} reduces the attack success rate to 10\% with only a 3.33\% drop in utility performance, outperforming state-of-the-art baselines.
\end{abstract}


\section{Introduction}
The widespread deployment of Large Language Models (LLMs) in mobile and edge applications has spawned Language Model-as-a-Service (LMaaS), with pre-trained LLMs cloud-hosted for users to submit task queries~\cite{sun2022black,thapa2022splitfed,openai_finetuning_guide,su2024titanic}. As task complexity grows, users increasingly require customized LLMs (\eg, personalized intelligent assistants) to cope with personalized scenarios~\cite{cai2024llmaas}. To meet this demand, service providers have introduced fine-tuning APIs that allow users to customize models on their private data. For example, OpenAI’s fine-tuning API enables users to upload domain-specific data (\eg, personal documents) in \texttt{JSON} format for tailored model development~\cite{openai_finetuning_guide}. However, this approach requires uploading sensitive data to the cloud, creating a fundamental tension between performance and privacy. Although platforms like OpenAI claim to delete user data within 30 days~\cite{openai_finetuning_guide}, such assurances may not be sufficient to alleviate concerns among privacy-sensitive users. This privacy concern is exacerbated in scenarios where data is inherently sensitive and regulated, \eg, in healthcare~\cite{arora2023promise} and finance~\cite{xu2024fwdllm}.

To alleviate such privacy concerns, a straightforward approach is the on-device adaptation technique~\cite{zhuang2024litemoe,ding2023parameter,gao2025federated}. The key idea is to allow user clients to download LLMs from the cloud and perform local fine-tuning by updating only a small set of trainable parameters, \ie, using Parameter-Efficient Fine-Tuning (PEFT)~\cite{zhuang2024litemoe,yu2024edge,kuang2024federatedscope} methods such as Low-Rank Adaptation (LoRA)~\cite{hulora}, without uploading sensitive data. However, these approaches typically assume full access to the cloud LLM weights on the device, and thus such deployment may violate model ownership restrictions ~\cite{zhang2025secret,wang2025never,duan2023flocks,wangflora}.

In light of the above challenges, device-cloud collaboration techniques are developed to achieve secure and efficient fine-tuning. These techniques propose offloading part of the fine-tuning task to the device, enabling the cloud and device to collaboratively complete fine-tuning of LLMs~\cite{xiao2023offsite,yao2025scaleot,wu2024fedbiot,yao2024split,lin2024splitlora,li2023privacy}. Specifically, the cloud offloads a portion of the LLM (\eg, the embedding layers)~\cite{yao2024split,lin2024splitlora,shen2025sap} or a compressed version of the model~\cite{xiao2023offsite,yao2025scaleot,wu2024fedbiot}, to the user device. The user then uses this component to generate intermediate representations (\eg, embeddings or model updates), which are shared with the cloud for joint fine-tuning. To further protect the privacy of these representations, one naive way is to instruct devices to inject Differential Privacy (DP)~\cite{dwork2006differential} noise before transmission, mitigating the risk of privacy inference attacks (\eg, embedding inversion attacks)~\cite{zhu2019deep,qu2025prompt,song2020information,huang2024transferable}. In this way, device-cloud collaboration avoids exposing the full cloud LLM to edge devices, reduces local resource demands, and keeps raw data on the device. However, this paradigm has a critical bottleneck in practice: \textit{the privacy-utility trade-off is difficult to achieve due to the injection of costly DP noise.}

In this paper, we propose \texttt{PrivTune}, a device-cloud collaboration framework for efficient and privacy-preserving fine-tuning in LMaaS. \texttt{PrivTune} achieves privacy protection for users while maintaining high model utility. It leverages Split Learning (SL) by offloading a tiny portion of the cloud LLM, \ie, the embedding layers (called the \textit{bottom model}), to the user device, where it extracts local representations. Fine-tuning is then performed efficiently on the cloud using a LoRA module, with only perturbed intermediate representations transmitted from the device, minimizing privacy leakage and computational burden on the users. Within this framework, the key idea is to add carefully crafted noise to the intermediate representations produced by the on-device bottom model. However, only slight noises can be tolerated to preserve model performance, which presents a key challenge. 

To this end, we formulate the noise design as an optimization problem that jointly considers both defense and utility objectives. Specifically, we aim to minimize the similarity between semantically related $n$-hop indirect neighbor tokens and maximize the intra-class cluster distance. This disrupts structured token distributions and enhances resistance to privacy inference attacks. Building atop, we further constrain the $\ell_2$-norm and distribution drift of embeddings before and after perturbation to preserve model utility. However, we notice that while the optimization formulation yields an optimal noise vector, it is deterministic. This makes it vulnerable to reverse engineering by an attacker, as the privacy guarantee fundamentally relies on noise being a random perturbation rather than a predictable signal~\cite{dwork2006differential,pan2020privacy,feyisetan2020privacy}. To resolve this, we design a token importance-aware ${d_\chi}$-Privacy mechanism. It uses the optimal noise vector to set the mean of the noise distribution, ensuring that the expected noise direction aligns with the optimization goal and meets randomness. Finally, we compute token importance scores using statistical features and attention weights, and use them to guide weighted noise injection, \ie, adding more noise to less important tokens while protecting critical semantic information.

We conduct extensive experiments on five benchmark datasets, covering both classification and generation tasks, and evaluate \texttt{PrivTune} across LLMs of varying sizes under three embedding inversion attacks and three attribute inference attacks. Specifically, we compare \texttt{PrivTune} with five state-of-the-art (SOTA) baselines, \ie, DP-Forward~\cite{du2023dp}, SLFT-DP~\cite{WANG2024Selective}, SAP~\cite{shen2025sap}, ScaleOT~\cite{yao2025scaleot}, and FedBiOT~\cite{wu2024fedbiot}. Our results demonstrate that \texttt{PrivTune} effectively mitigates these privacy attacks while preserving model utility. Specifically, \texttt{PrivTune} reduces the success rate of embedding inversion attacks to approximately 10\%, with only a 3.33\% drop in utility performance, surpassing all existing SOTA baselines. Moreover, experimental results show that \texttt{PrivTune} is efficient in terms of computational and communication overhead, highlighting its effectiveness in real-world scenarios.

Our contributions can be summarized as follows:
\begin{icompact}
	\item We propose \texttt{PrivTune}, an efficient and privacy-preserving fine-tuning framework, to mitigate inference attacks on language models with a formal utility guarantee.
	
	\item We formulate a defense-utility optimization problem to craft $d_\chi$-Privacy noise with token importance that disrupts the correlation between embedding vectors, effectively defending against privacy attacks.

	\item We conduct comprehensive experiments to demonstrate that our \texttt{PrivTune} effectively defends against six privacy inference attacks and outperforms all existing methods in terms of the defense-utility trade-off and overhead.
\end{icompact}

\section{Related Work}\label{sec:related}
\para{Parameter-Efficient Fine-Tuning.} PEFT methods have emerged as a practical solution to reduce the computational and memory overhead of adapting LLMs~\cite{hulora,ding2023parameter,lester2021power} on device. Instead of updating the entire model, PEFT focuses on fine-tuning a small subset of parameters, enabling faster and more resource-efficient adaptation. For example, LoRA~\cite{hulora} introduces low-rank matrices into existing weight matrices, significantly reducing the number of parameters that need to be updated. Despite their efficiency, most PEFT approaches assume full access to the LLM and lack built-in privacy protections, making them unsuitable for sensitive on-device applications where user data privacy must be preserved~\cite{xu2024fwdllm}.

\para{Device–Cloud Collaboration for Fine-Tuning.} 
The growing demand for data privacy and personalization has driven research on device–cloud collaboration for fine-tuning of LLMs. Existing research has made preliminary progress along two main directions: Offsite-Tuning~\cite{xiao2023offsite,yao2025scaleot,wu2024fedbiot} and Split Learning-based Fine-Tuning (SLFT)~\cite{yao2024split,lin2024splitlora,li2023privacy}. To start with, Offsite-Tuning avoids sharing the full cloud LLM to edge devices. Instead, it transmits a compressed version of the model, referred to as an \textit{Emulator}, using model compression~\cite{xiao2023offsite} or knowledge distillation~\cite{yao2025scaleot} techniques, enabling local fine-tuning via on-device adapters (\eg, LoRA). While this approach mitigates privacy concerns for users (as now the fine-tuning process will happen locally on the users' devices), it still poses risks to model ownership due to the structural similarity between the \textit{Emulator} and the original cloud LLM~\cite{wu2024fedbiot,shen2025sap}. In contrast, SLFT offloads part of the cloud LLM (called \textit{the bottom model}) to the edge device and enables privacy-preserving fine-tuning by exchanging differentially private intermediate embeddings~\cite{li2023privacy}. While this makes it harder for the edge device to reconstruct the full model, the added noise from DP~\cite{dwork2006differential} often leads to a poor privacy-utility trade-off.

\section{Background}\label{sec:related:prel}
\para{Split Learning-based Fine-Tuning.}
We consider an SLFT scenario where a service provider $\mathcal{S}$ holds an LLM parameterized by $\bm{w}$, and offers customized fine-tuning services to users $\mathcal{U}$, where each user $u_i$ possesses a private dataset $\mathcal{D}_i$ containing sensitive textual data. To enable secure and resource-efficient fine-tuning, the LLM is partitioned into two components via SL~\cite{thapa2022splitfed,romanini2021pyvertical}: a \textit{bottom model} $f_{\theta_c}$ and a \textit{top model} $g_{\theta_s}$, such that the full model is represented as: $\mathcal{F}(x; \bm{w}) = g_{\theta_s}(f_{\theta_c}(x))$, where $f_{\theta_c}$ is deployed locally on the user's edge device, $g_{\theta_s}$ is retained on the server, $\bm{w} = \theta_c \cup \theta_s$ denotes the overall model parameters. The forward pass proceeds as follows. Given an input $x \in \mathcal{D}_i$, the device computes an intermediate representation: $\bm{h} = f_{\theta_c}(x),$ and transmits $\bm{h}$ to the server. The server then computes the final prediction: $\hat{y} = g_{\theta_s}(\bm{h}).$ Since the label $y$ remains private and is only available on the user side, the backward pass is collaboratively performed. The above process can be formulated as:
\begin{equation}   \label{eqn2}                     
	\arg \min_{\bm{\delta}} \mathcal{L}(\bm{w}+\bm{\delta}, \mathcal{D}_i),
\end{equation}
where $\bm{\delta}$ is the trainable parameters of the PEFT methods, \eg, LoRA~\cite{hulora}. This training process proceeds iteratively until fine-tuned model convergence. To prevent SLFT from privacy inference attacks~\cite{du2023dp,song2020information,shokri2017membership,huang2024transferable} (detailed in the following sections), existing approaches typically rely on DP noise mechanisms to provide formal privacy guarantees. However, such methods often incur significant performance degradation. This highlights the need for slight perturbation-based defenses that can offer privacy protection with minimal impact on utility, as explored in later sections (see \S\ref{method}).

\para{Embedding Inversion Attacks (EIAs).}
EIAs aim to reconstruct input tokens $x$ from intermediate representations or gradients exposed during training. These attacks exploit the nearly injective mapping of early-layer representations, particularly the input embedding layer $\mathcal{E}: \mathcal{V} \to \mathbb{R}^d$, where $\mathcal{V}$ is the vocabulary. Several classic attacks are defined as follows:

\textit{Attack-0: Direct Activation Inversion~\cite{song2020information}.} Given the activation $\bm{h} = \phi_d(x)$, an attacker attempts to find $\hat{x}$ such that:
\begin{equation}
\hat{x} = \arg\min_{x'} \| \phi_d(x') - \bm{h} \|_2^2,
\end{equation}
where $\phi_d(\cdot)$ is the embedding model hold by the attacker, and $\|\|_2^2$ is the $\ell_2$ distance. %In transformer models, where $\phi_d$ is often a stack of self-attention and feedforward layers, this inversion is nontrivial but feasible with access to similar model parameters.

\textit{Attack-1: Gradient Inversion Attack~\cite{zhu2019deep}.} When gradients $g=\nabla_{\theta_c} \mathcal{L}(f_{\theta_c}(x), y)$ are visible to an attacker (\eg, via model updates), the attack recovers input $x$ by minimizing:
\begin{equation}
\hat{x} = \arg\min_{x'} \left\| \nabla_{\theta_c} \mathcal{L}(\phi(x'; \theta), y) - g \right\|_2^2,   
\end{equation}
where $g$ is the observed gradient.

\textit{Attack-2: Embedding Matching and Nearest Neighbor Recovery~\cite{huang2024transferable}.} In some cases, attackers can simply compute cosine similarity between leaked embeddings $\bm{h}_i$ and all entries in the vocabulary embedding matrix $\mathcal{E} \in \mathbb{R}^{|\mathcal{V}| \times d}$:
\begin{equation}
\hat{t}_i = \arg\max_{t \in \mathcal{V}} \cos(\bm{h}_i, \mathcal{E}),   
\end{equation}
recovering the most probable input token $t_i$ corresponding to $\bm{h}_i$. Even if raw data $x$ remains local, the intermediate features $\bm{h}$ can leak semantic content with high fidelity.

\para{Attribute Inference Attacks (AIAs).} In addition to reconstructing $x$, an attacker may aim to infer sensitive attributes from a user's data, \eg, gender, location, or political affiliation, through AIAs. Although the raw input $x$ remains local, the intermediate representation $\bm{h} = f_{\theta_c}(x; \theta_c)$ may still encode enough semantic information to allow attribute inference. A typical AIA assumes that a specific private attribute $a \in \mathcal{A}$ (\eg, ``is this input from a female user?'') is correlated with the distribution of activations $\bm{h}$. Given access to $\bm{h}$, an attacker attempts to learn a mapping $\mathcal{A}: h \rightarrow \hat{a}$ such that:
\begin{equation}
\hat{a} = \arg\max_{a' \in \mathcal{A}} \Pr(a' \mid \bm{h}).
\end{equation}
Several classes of AIA have been studied in SL settings:

\textit{Attack-3: Supervised Attribute Classifier~\cite{du2023dp}.} With access to a labeled shadow dataset, the attacker trains a dedicated classifier $f_{\theta}$  to predict private attributes directly from intermediate features. Specifically, the objective of the attacker is defined as follows:
\begin{equation}
p = f_{\theta}\left(\frac{1}{n} \sum_{i=1}^{n} \bm{h}_i\right),\mathcal{L}_{\text{AIA}} = -\sum_{i=1}^{|C|} t_i \log p[i],
\end{equation}
where is the private attribute $t \in \{0, 1\}^{|C|}$ from the hidden representations, $|C|$ is the number of private classes, and $f_{\theta}$ is a designed classifier (\ie, Multilayer Perceptron model). %a 2-layer MLP with 768 hidden units and a ReLU activation function, following the settings in \cite{plant2021cape}.

\textit{Attack-4: Gradient-based Attribute Inference~\cite{shokri2017membership}.}
In multi-round SL, when the server can observe gradients $\nabla_{\theta_s} \mathcal{L}$, the attacker hypothesizes that certain attributes disproportionately influence these updates. The attacker computes:
\begin{equation}
\hat{a} = \arg\max_{a'} \Pr(a' \mid \nabla_{\theta_S} \mathcal{L}),
\end{equation}
by training a meta-classifier or using attribution techniques.

\begin{table}[!t]
	\centering
	\caption{Six attacks and their background knowledge.}
	\label{tab:aia-eia-comparison}
	\adjustbox{width=0.45\textwidth}{%
		\begin{tabular}{lccccc}
			\toprule
			\textbf{Attack} &
			\makecell{\textbf{Requires} \\ \textbf{Model Access}} &
			\makecell{\textbf{Requires} \\ \textbf{Gradients}} &
			\makecell{\textbf{Requires} \\ \textbf{Embeddings}} &
			\makecell{\textbf{Requires} \\ \textbf{Labels}} &
			\makecell{\textbf{Supervised} \\ \textbf{Shadow Data}} \\
			\midrule
			\multicolumn{6}{l}{\textit{\textbf{Embedding Inversion Attacks (EIAs)}}} \\
			\midrule
			Attack-0 & \cmark & \xmark & \cmark & \xmark & \xmark \\
			Attack-1 & \cmark & \cmark & \xmark & \xmark & \xmark \\
			Attack-2 & \xmark & \xmark & \cmark & \xmark & \xmark \\
			\midrule
			\multicolumn{6}{l}{\textbf{\textit{Attribute Inference Attacks (AIAs)}}} \\
			\midrule
			Attack-3 & \cmark & \xmark & \cmark & \cmark & \cmark \\
			Attack-4 & \cmark & \cmark & \xmark & \cmark & \xmark \\
			Attack-5 & \cmark & \xmark & \cmark & \xmark & \xmark \\
			\bottomrule
	\end{tabular}}
\end{table}

\textit{Attack-5: Representation Clustering Attack~\cite{song2020information}.} The attacker applies unsupervised clustering (\eg, K-means) to the intermediate activations $\bm{h}$ collected during SL. After unsupervised clustering, the attacker maps cluster centroids to attribute classes using shadow-labeled data. The core assumption is:
\begin{equation}
  \mathbb{E}[\bm{h} \mid a_1] \ne \mathbb{E}[\bm{h} \mid a_2], \quad \forall a_1 \ne a_2 \in \mathcal{A}.  
\end{equation}
To facilitate the understanding of the above attacks, we summarize their background knowledge in Table \ref{tab:aia-eia-comparison}.
%To facilitate the understanding of the core assumptions of the above attacks, we summarize the background knowledge of the above attacks in Table \ref{tab:aia-eia-comparison}.

\section{Problem Formulation}\label{sec:method:system}
This section clarifies our threat model and defense objectives and then formulates our defense problem.
\subsection{Threat Model}
\para{Attacker’s Goal.} In this paper, we assume that the service provider is honest but curious; that is, it always follows the designed fine-tuned protocol but is curious about the private information of participants (\eg, users’ private data). Therefore, the service provider, acting as a potential adversary, may attempt to reconstruct or obtain the user's private data or attributes through inference attacks (\ie, EIAs and AIAs). %In the context of privacy-preserving fine-tuning services, such an attack is deemed successful if the service provider is able to infer the user's private data or attributes by any means of inference.

\para{Attacker’s Background Knowledge.} Since our work focuses on developing defenses, we assume a strong adversarial model where the attacker possesses significant background knowledge. Specifically, the attacker is assumed to have access to the intermediate representations transmitted by the user, as well as full knowledge of the user's bottom model, including its parameters, architecture, and version, \ie, a white-box setting. Additionally, the attacker is assumed to have white-box access to the perturbed intermediate representations produced by our defense mechanism. However, the attacker does not know the specific perturbation technique applied, as it is carefully designed and kept confidential by the defender~\cite{lou2025grid}. %This assumption is realistic and consistent with standard practices in security, such as in cryptographic systems, where the secrecy of the key is a fundamental part of the threat model~\cite{lou2025grid}.

 %a strong adversary with full access to the edge device's bottom model, including its architecture, parameters, and version, as well as the intermediate representations generated during the fine-tuning process. In particular, 

\para{Attacker's Capabilities.} We assume the attacker is capable of arbitrarily manipulating these intermediate representations. It then uses them to infer private information by adopting the state-of-the-art embedding inversion attacks \cite{du2023dp} and attribute inference attacks \cite{song2020information}. However, the attacker cannot alter the design or execution of the fine-tuning service protocol or the defense mechanisms employed by the edge device. %For instance, the on-device defense technique is executed locally in a confidential manner and remains inaccessible to the attacker, preventing any direct manipulation.

\subsection{Defense Formulation}
Given the above threat model, aligned with the previous work~\cite{wang2023privatelora,qu2021natural}, our goal is to achieve: 
1) the service provider cannot recover the original input text or infer private attributes from the transmitted text representation; 
2) the proposed defense should maintain comparable performance compared to non-private methods. Next, we detail our motivation and provide mathematical formulations for our defense methods.

\para{Our Motivations.} Although various privacy inference attacks on SLFT differ in their methods, they often rely on the same threat model: the intermediate representations (\ie, embeddings) exchanged between the device and cloud can leak private input or label information. Therefore, an effective defense should focus on perturbing these representations to resist both EIA and AIA attacks. However, these intermediate representations are based on important semantic features needed for model performance. Adding perturbations during fine-tuning may harm utility. To address this, we propose adding controlled noise to the transmitted representations after tokenization. This reduces their semantic consistency and similarity, hiding attackable patterns while preserving utility. Formally, let $\bm{h}$ denote the clean intermediate output, and $\bm{p}$ be the protective noise, the transmitted representation becomes:
\begin{equation}
	\bm{\tilde h} = \bm{h} + \bm{p}.
\end{equation}
In doing so, we make it significantly harder for the attacker to invert or infer private information from intermediate outputs~\cite{huang2024transferable,abadi2016deep}. Additionally, we will formulate this noise generation as an optimization problem to ensure that both defense goals and model utility are maintained.

\para{Defense Goal.} To defend against inference attacks that exploit semantic or structural similarity among intermediate representations, our method aims to inject perturbations into the activations before transmitting them to the cloud. Specifically, the goal is to obfuscate patterns that attackers rely on when performing embedding inversion or activation inference attacks while ensuring that the resulting representations still support high model performance.

For EIA, the attack's success relies on the structural consistency of embeddings across different inputs. To disrupt this, we propose minimizing the similarity between embedding vectors from semantically distant inputs. For instance, if $\bm{h}_i$ and $\bm{h}_j$ are the intermediate embeddings of two inputs $x_i$ and $x_j$, and their labels differ ($y_i \neq y_j$), then we compute their similarity:
\begin{equation}\label{eq-11}
\text{sim}(\bm{h}_i, \bm{h}_j) = \text{corr}(\bm{h}_i, \bm{h}_j) + \cos(\bm{h}_i, \bm{h}_j),  
\end{equation}
where $\text{corr}(\cdot)$ and $\cos(\cdot)$ measure the Pearson correlation and cosine similarity, respectively. %High similarity among unrelated inputs provides exploitable cues for inversion, hence minimizing this similarity increases the ambiguity in the embedding space.

For AIA, the attacker targets semantic clustering in the latent space to infer class labels. In this case, we aim to prevent strong intra-class compactness in the representations. That is, for embeddings from the same class ($y_i = y_j$), we introduce controlled dispersion to reduce label inferability. This is achieved by maximizing the distance or dissimilarity between embeddings from the same class to prevent tight clustering. Putting these two objectives together, we formulate the noise vector generation as an optimization problem:
\begin{equation*}\label{eq-12}
{\text{\textbf{OPT-1:}}}\quad {\min}\underbrace {\sum\limits_{{y_i} \ne {y_j}} {sim(} {{\widetilde {\bm{h}}}_{\text{i}}},{{\widetilde {\bm{h}}}_{\text{j}}})}_{{\text{Resist EIA}}} - \underbrace {\lambda \sum\limits_{{y_i} = {y_j}} {\|} {{\widetilde {\bm{h}}}_{\text{i}}}{\text{ - }}{{\widetilde {\bm{h}}}_{\text{j}}}\|_2^2}_{{\text{Resist AIA}}},
\end{equation*}
where $\lambda$ is a tunable hyperparameter. Note that in the generation task, the label $y$ can be predefined by the user based on keywords or semantics in the local data. %The first term reduces embedding consistency across distinct inputs to hinder inversion attacks, while the second term introduces controlled variation among same-class representations to mitigate activation inference.Let $\bm{h} = f_{\theta_c}(x)$ denote the intermediate representation generated by the on-device bottom model, and let $\tilde{\bm{h}} = \bm{h} + \bm{p}$ denote the perturbed version.
\begin{figure*}[!t]
	\centering
	\includegraphics[width=1\textwidth]{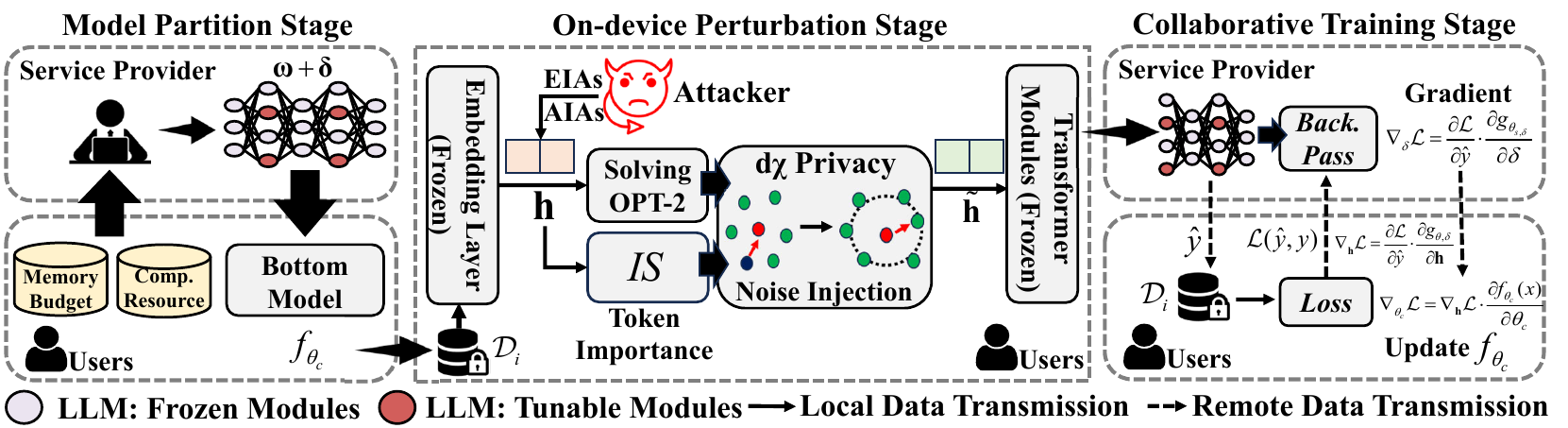}
	\caption{
		Overview of the \texttt{PrivTune} framework, where the cloud LLM is split into a bottom model and a top model.
	}
	\label{fig-1}
\end{figure*}

%The user holds the bottom model and privatizes the embedding representations before sending them to the server during the forward process.
%Upon receiving the embedding representations, the server computes the loss and starts the backward process.

\para{Utility Goal.} To ensure that the performance of the target model is not compromised, we introduce three constraints that collectively preserve the model's predictive utility across both classification and generation tasks, as detailed below.

\textit{Prediction Consistency.} For classification tasks, we enforce that the predicted class remains unchanged after perturbation:
\begin{equation}\label{eq-13}
\arg\max_{a \in \{1, \dots, C\}} \left[g_{\theta_s}(\bm{h})\right]^a = \arg\max_{a \in \{1, \dots, C\}} \left[g_{\theta_s}(\tilde{\bm{h}})\right]^a,
\end{equation}
where $g_{\theta_s}(\cdot)$ denotes the output logits of the server-side model and $a$ indexes the predicted class among $C$ categories.

For generation tasks, we require that the semantic content of the output text remain similar before and after perturbation. Let $y = \text{Dec}(g_{\theta_s}(\bm{h}))$ and $\tilde{y} = \text{Dec}(g_{\theta_s}(\tilde{\bm{h}}))$ be the decoded sequences. We enforce:
\begin{equation}\label{eq-14}
\text{Sim}(y, \tilde{y}) \geq \tau,
\vspace{-0.2cm}
\end{equation}
where $\text{Sim}(\cdot, \cdot)$ is a task-agnostic similarity metric (\eg, ROUGE-L), and $\tau \in [0,1]$ is a predefined threshold.
% (\eg, BLEU, ROUGE, or embedding-based cosine similarity)

\textit{Semantic Preservation.} We constrain the angular deviation between the original and perturbed embeddings to ensure semantic information is preserved~\cite{qiu2025flm}:
\begin{equation}\label{eq-15}
\cos(\bm{h}, \tilde{\bm{h}}) \geq \delta,
\end{equation}
where $\delta > 0$ is the cosine similarity threshold. %This ensures that the directional structure of the embedding vector remains largely unchanged.

\textit{Distributional Validity.} To prevent $\tilde{\bm{h}}$ from drifting out of the training data distribution, we constrain its deviation from the empirical support of the embedding space:
\begin{equation}\label{eq-16}
\| \tilde{\bm{h}} - \boldsymbol{\mu} \|_2 \leq R,
\end{equation}
where $\boldsymbol{\mu}$ is the mean of training embeddings and $R$ is the maximum observed radius of the original embedding space. 

We now unify our defense goal with the three utility-preserving constraints into a joint objective:
\begin{equation*}\label{eq-17}
   \begin{aligned}
\textbf{OPT-2:} \quad
&\min \; \sum_{y_i \neq y_j} \text{sim}(\tilde{\bm{h}}_i, \tilde{\bm{h}}_j) \;-\; \lambda \sum_{ y_i = y_j} \| \tilde{\bm{h}}_i - \tilde{\bm{h}}_j \|_2^2 \\
&s.t. \quad  \text{\eqref{eq-11}, \eqref{eq-13}--\eqref{eq-14}, \eqref{eq-15},  \text{and} \eqref{eq-16}}.
\end{aligned} 
\end{equation*}
This formulation aims to find the optimal noise vector to achieve the defense-utility goal.%The embedding vector $\bm{h}$ in the formulation is a constant and is generated by the bottom model. The noise vector $\bm{p}$ is the variable to be solved. 

\section{\texttt{PrivTune}}
\subsection{Overview}
In this paper, we propose \texttt{PrivTune}, a device-cloud collaboration framework built upon the SplitNN architecture (the classic SL configuration)~\cite{thapa2022splitfed,romanini2021pyvertical} to enable privacy-preserving and efficient fine-tuning of LLMs, as illustrated in Fig. \ref{fig-1}. The framework consists of three key stages: model partition, on-device perturbation, and collaborative training. Next, we briefly introduce the execution details of each stage.

\para{Model Partition Stage.} In this stage, the service provider determines the maximum number of bottom model layers $m_{\text{max}}$ that can be deployed on the user's edge device, based on its resource profile (\eg, memory, RAM, and compute capability). The provider then transmits the first $m$ layers ($m < m_{\text{max}}$) of an $l$-layer LLM to the user, which includes the embedding layer and several encoder blocks (\ie, Transformer modules). The remaining $l - m$ layers, comprising the higher-level encoder and output layers, are retained as the top model by the service provider.

\para{On-device Perturbation Stage.} At this stage, the edge device performs a forward pass through its local embedding layers to generate intermediate token representations. To reduce computational overhead on resource-constrained devices, we adopt the post-fine-tuning approach~\cite{sun2022black}, wherein the embedding layers and Transformer modules of the bottom model are frozen during fine-tuning. We first compute an optimal noise vector by solving \textbf{OPT-2} (see \S\ref{noise}). This vector serves as a directional guide for token importance-aware ${d_\chi}$-Privacy noise sampling (see \S\ref{method}), where the sampled noise is then added to the intermediate representation to protect privacy while preserving model utility.

\para{Collaborative Training Stage.} Finally, the edge device transmits the perturbed representation to the service provider, who completes the forward pass using the top model. Since the labels remain on the edge device, collaborative training is necessary to update the trainable parameters (\eg, those in a LoRA module). The service provider sends the model output back to the user, who then computes the gradient of the loss with respect to the output. This gradient is returned to the service provider to update the trainable parameters. The above process is repeated iteratively until the fine-tuning converges.

%\vspace{-0.5cm}
\subsection{Key Challenges and Our Intuitions}
\para{Key Challenges.} We identify three key challenges when solving the optimization problem \textbf{OPT-2} in practice:
\begin{icompact}
\item \textbf{C1: Efficient and Scalable Optimization.}
Solving \textbf{OPT-2} over a large-scale dataset with pairwise similarity constraints is computationally expensive. For each training step, computing similarities or distances between embedding pairs (both intra- and inter-class) scales quadratically with the dataset size (\ie, $\mathcal{O}(N^2)$). Efficient approximations are necessary for practical deployment.%Moreover, constraints like out-of-distribution bounding (Eq. \eqref{eq-16}) require maintaining empirical statistics over the entire representation space, which may be costly to compute and update. 

\item \textbf{C2: Balancing Noise Strength and Utility Constraints.} Injecting strong noise helps reduce representational similarity (to defend against EIA and AIA), but also increases the risk of violating utility constraints, \eg, \eqref{eq-13}--\eqref{eq-14} and \eqref{eq-15}. Since these objectives are inherently conflicting, naive noise injection often leads to performance collapse or insufficient privacy. Designing an adaptive mechanism that tailors the noise scale to both the token’s role and the representation's context is therefore critical.
%(to defend against EIA and AIA)
\item \textbf{C3: Token-wise Importance Estimation.} The optimization relies on identifying which tokens are semantically or structurally sensitive. However, determining token importance dynamically across diverse inputs is non-trivial. Classical metrics like attention weights or gradient saliency maps can be noisy or unstable, especially in low-resource or fine-tuning scenarios. Moreover, directly measuring semantic sensitivity \eqref{eq-13}--\eqref{eq-14} without access to the full model (especially the top model) introduces additional uncertainty.

\end{icompact}

\para{Our Intuitions.} To address the above challenges, we propose the following intuitions to design our approach:
\begin{icompact}
\item To address the \textbf{C1}, we decouple the pairwise constraints into nearest neighbor disguise constraints and utilize triangle inequality to estimate support radius and distribution center. This enables the computations and constraints in \textbf{OPT-2} to be easily performed on resource-constrained devices.

\item To address the \textbf{C2}, we leverage the solution of \textbf{OPT-2} to guide the configuration of the noise scale parameter in the ${d_\chi}$-Privacy. Specifically, we derive the ideal perturbation vector as the mean of the Laplace noise distribution within the ${d_\chi}$-Privacy mechanism, incorporating utility goals.

%we use a learned mapping from token importances to ${d_\chi}$-Privacy noise scale combined with utility goals whose upper and lower bounds are derived from task-specific tolerances.

\item To address the \textbf{C3}, we leverage both word frequency features and entropy-weighted attention aggregation to construct a token importance evaluation module. Based on the resulting importance scores, tokens are then adaptively perturbed with noise at varying levels.

\end{icompact}

\subsection{\texttt{PrivTune} Solution}\label{noise}
Following our intuitions, we rectify the optimization
problem of \textbf{OPT-2} in calculating the noise vectors.

\subsubsection{Reformulating \textbf{OPT-1}} 
To reduce the computational cost of defense optimization, we propose a reformulated version of \textbf{OPT-1} that avoids exhaustive pairwise comparisons across all token embeddings. Specifically, we implement \textit{token disguise} by minimizing the similarity gap between a token’s $k$-nearest neighbors and its $n$-hop indirect neighbors, rather than comparing all token pairs. Let $x_i$ be a token, with its $k$-nearest neighbor token set denoted by $\mathcal{P}_i$, and its $n$-hop indirect neighbor token set denoted by $\mathcal{Q}_i$. Given a noise vector $\bm{p}_i$ applied to $x_i$, the similarity gap $D_i$ is computed as:
\begin{equation}
\footnotesize
D_i^{{EIA}} =  \frac{1}{|\mathcal{P}_i|} \sum_{x_j \in \mathcal{P}_i} \text{sim}(\tilde{\bm{h}}_i, \bm{h}_j) - \frac{1}{|\mathcal{Q}_i|} \sum_{x_k \in \mathcal{Q}_i} \text{sim}(\tilde{\bm{h}}_i, \bm{h}_k).
\end{equation}

The AIA defense component avoids overly compact clusters by limiting how close a token gets to its own class centroid:
\begin{equation}
D_i^{AIA} = \lambda \cdot \| \tilde{\bm{h}}_i - \boldsymbol{\mu}_{y_i} \|_2^2,
\end{equation}
where $\boldsymbol{\mu}_{y_i} = \frac{1}{|\mathcal{C}_{y_i}|} \sum_{j \in \mathcal{C}_{y_i}} \tilde{\bm{h}}_j$ is the class centroid and $\mathcal{C}_{y_i}$ is the set of tokens in the same class $y_i$. Combining the two objectives, the final reformulated \textbf{OPT-1} becomes:
\begin{equation}
\min \sum_i \left( D_i^{{EIA}} - D_i^{{AIA}} \right).
\end{equation}
Both $\mathcal{P}_i$ and $\mathcal{Q}_i$ can be efficiently precomputed using either frozen embedding layers or structural proxies (\eg, attention graphs or word co-occurrence networks). The pseudo-label $y_i$ can be inferred from lightweight proxy classifiers or semantic clustering, eliminating the need for access to the cloud model.

\para{Remark.} In the original \textbf{OPT-1}, both the EIA and AIA objectives involve quadratic $\mathcal{O}(N^2)$ complexity due to pairwise comparisons across tokens. The reformulated version significantly reduces this cost. For EIA, the use of local $k$-nearest and $n$-hop neighbors lowers the complexity to $\mathcal{O}(Nk)$, where $k \ll N$. For AIA, replacing all intra-class pairwise distances with the distance to a precomputed class centroid reduces its complexity from $\mathcal{O}(N^2)$ to $\mathcal{O}(N)$. This reformulation enables scalable optimization for large datasets.

\subsubsection{Reformulating Constraints}
Given that the cloud model $g_{\theta_s}$ is inaccessible and the on-device embeddings are norm-bounded (\ie, $|{\bm{h}}|_2 \leq B$), we reformulate the original utility constraints into a compact and tractable form. The prediction consistency \eqref{eq-13}--\eqref{eq-14} and semantic preservation \eqref{eq-15} objectives are jointly approximated by a local proximity constraint that enforces small $\ell_2$ deviation between the original embedding $\bm{h}$ and the perturbed embedding $\tilde{\bm{h}}$:
\begin{equation}
\| \tilde{\bm{h}} - \bm{h} \|_2^2 \leq 2{B^2}(1 - \delta ).
\end{equation}
This constraint implicitly preserves both model prediction behavior (assuming local Lipschitz continuity of $g_{\theta_s}$~\cite{hulora}) and the semantic integrity of the input. 

\begin{algorithm}[!t]
	\small
	\caption{Projected Gradient Descent for \textbf{OPT-3}}\label{algo-1}
	\begin{algorithmic}[1]
		\REQUIRE Initial embeddings $\{\bm{h}_i\}$, centroids $\{\boldsymbol{\mu}_{y_i}, \boldsymbol{\mu}\}$, learning rate $\eta$, max iterations $T$
		\STATE Initialize $\bm{p}_i^{(0)} \gets \bm{0}$ for all $i$
		\FOR{$t = 0$ to $T-1$}
		\FOR{each token $x_i$}
		\STATE Compute $\tilde{\bm{h}}_i^{(t)} \gets \bm{h}_i + \bm{p}_i^{(t)}$
		\STATE Compute gradient: $\nabla \gets \nabla_{\bm{p}_i} (D_i^{EIA} - D_i^{AIA})$
		\STATE Update: $\bm{p}_i^{(t+1)} \gets \bm{p}_i^{(t)} - \eta \cdot \nabla$
		\STATE $\tilde{\bm{h}}_i^{(t+1)} \gets \bm{h}_i + \bm{p}_i^{(t+1)}$
		\IF{$\| \tilde{\bm{h}}_i^{(t+1)} - \bm{h}_i \|_2^2 > 2B^2(1 - \delta)$}
		\STATE \textcolor{gray}{\% Project onto L1:} \\$\tilde{\bm{h}}_i^{(t+1)} \gets \bm{h}_i + \frac{2B\sqrt{1 - \delta}}{\| \tilde{\bm{h}}_i^{(t+1)} - \bm{h}_i \|_2} (\tilde{\bm{h}}_i^{(t+1)} - \bm{h}_i)$ 
		\ENDIF
		\IF{$\| \tilde{\bm{h}}_i^{(t+1)} - \boldsymbol{\mu} \|_2 > R$}
		\STATE \textcolor{gray}{\% Project onto L2:}\\ $\tilde{\bm{h}}_i^{(t+1)} \gets \boldsymbol{\mu} + \frac{R}{\| \tilde{\bm{h}}_i^{(t+1)} - \boldsymbol{\mu} \|_2} (\tilde{\bm{h}}_i^{(t+1)} - \boldsymbol{\mu})$
		\ENDIF
		\STATE Update noise vector: $\bm{p}_i^{(t+1)} \gets \tilde{\bm{h}}_i^{(t+1)} - \bm{h}_i$
		\ENDFOR
		\ENDFOR
		\RETURN $\{\bm{p}_i^{(T)}\}$ for all $i$
	\end{algorithmic}
\end{algorithm}

\subsubsection{Reformulation of OPT-2} Finally, we summarize all efforts to reformulate \textbf{OPT-2} into \textbf{OPT-3}. Given the embedding vector extracted by the bottom model on the device side, for each token $x_i\in \mathcal{D}_i$, we calculate its noise vector $\bm{p}_i$ by solving the following optimization problem.
\begin{equation*}\label{eq:opt2-reformulated-aligned}
	\begin{aligned}
		\textbf{OPT-3:} \quad
		\mathop {\min }\limits_{{{\bm{p}}_i}}  \quad &\sum_i \left( D_i^{\mathrm{EIA}}(\bm{p}_i) - D_i^{\mathrm{AIA}}(\bm{p}_i) \right), \\
		\;\; s.t. \quad &\text{L1: } \| \tilde{\bm{h}} - \bm{h} \|_2^2 \leq 2{B^2}(1 - \delta ), \forall i,\\
		&\text{L2: } \| \tilde{\bm{h}} - \boldsymbol{\mu} \|_2 \leq R, \forall i.
	\end{aligned}
\end{equation*}

\para{OPT-3 Solution.} To efficiently solve \textbf{OPT-3}, we employ Projected Gradient Descent (PGD)~\cite{madry2018towards} with dual projection steps to enforce the two convex constraints. The optimization is performed over the perturbation vector $\bm{p}_i$, where the perturbed embedding is defined as $\tilde{\bm{h}}_i = \bm{h}_i + \bm{p}_i$. At each iteration, we compute the gradient of the objective $\mathcal{L} = D_i^{\mathrm{EIA}} - D_i^{\mathrm{AIA}}$ with respect to $\bm{p}_i$, perform a gradient descent update, and project the result back onto the feasible region defined by the local perturbation constraint (L1) and the global distributional constraint (L2). Specifically, if the perturbation exceeds the local bound $\| \tilde{\bm{h}}_i - \bm{h}_i \|_2^2 > 2B^2(1 - \delta)$, we scale it back to lie on the boundary of the $\ell_2$-ball centered at $\bm{h}_i$; similarly, if $\| \tilde{\bm{h}}_i - \boldsymbol{\mu} \|_2 > R$, we project $\tilde{\bm{h}}_i$ back toward the global centroid $\boldsymbol{\mu}$ with radius $R$. We denote the optimal noise vector solved as $\bm{p}_i^*$. The procedure is efficient, parallelizable, and well-suited for on-device deployment, with convergence typically achieved in a small number of iterations when the learning rate is properly tuned. The details can be found in Algo. \ref{algo-1}.

\subsection{Token Importance-aware $d\chi$-Privacy}\label{method}  
To provide formal privacy guarantees, we adopt the ${d_\chi}$-Privacy mechanism~\cite{yang2022k} instead of conventional DP. The ${d_\chi}$-Privacy framework defines a smooth sensitivity bound tailored to continuous spaces such as embedding spaces, making it especially suitable for LLM representation perturbation.

\begin{definition}
	($d_\chi$-Privacy). A randomized mechanism $\mathcal{M}$ satisfies $d_\chi$-Privacy with a privacy budget $\epsilon > 0$ if for any two inputs $x, x' \in \mathcal{X}$ and any output $y \in \mathcal{Y}$, the following holds:
	\begin{equation}\label{eq-18}
		\ln \frac{{\Pr [\mathcal{M}(x) = y]}}{{\Pr [\mathcal{M}(x') = y]}} \leqslant \epsilon \cdot  {d_\chi }(x,x'),
	\end{equation}
	where ${d_\chi }( \cdot , \cdot )$ denotes the $\ell_2$ norm in the input space. 
\end{definition}
%This formulation generalizes traditional DP to continuous metric spaces by bounding the change in output distributions based on input distance. 

To instantiate this in \texttt{PrivTune}, we perturb token embeddings using noise calibrated to the sensitivity of the embedding function $f_{\theta_c}(\cdot)$ with respect to input perturbations. Specifically, we define the $\ell_2$-sensitivity as: $\Delta_\chi = \sup_{x, x'} \frac{\|f_{\theta_c}(x) - f_{\theta_c}(x')\|_2}{d_\chi(x, x')}$, and add noise accordingly:
\vspace{-0.3cm}
\begin{equation} \label{eqn3}
	\bm{p}_i \sim \text{Lap}_d(\bm{p}) \propto \exp\left(-\frac{\epsilon}{\Delta_\chi} |\bm{p}|_2\right).
\end{equation}	
This mechanism satisfies $d_\chi$-Privacy by construction, ensuring that the magnitude of perturbation is proportional to the maximum sensitivity of the embedding function with respect to input changes. To enhance the trade-off between utility and privacy, we design a token-level importance-aware ${d_\chi}$-Privacy mechanism. The key idea is to assign adaptive noise levels based on each token's importance score, \ie, less important tokens receive stronger perturbations while important ones are preserved to retain utility.

\para{Token Importance for Classification.} For classification tasks, we use the statistical features of tokens to represent their importance. We adopt a TF-IDF-inspired metric~\cite{aizawa2003information} to estimate the importance of token $x_i$ in a classification context. Let $p(x=x_i|y=c)$ be the frequency of token $x_i$ in class $c$, then the importance score ($IS$) of token $x_i$ is given by:
\begin{equation}          
	IS({x_i}) = \frac{1}{{|\mathcal{C}| - 1}}\sum\limits_{{c^\prime },{c^\prime } \ne c} {\ln \frac{{p(x = {x_i}|y = c)}}{{p(x = {x_i}|y = {c^\prime })}}} ,
\end{equation}
where the log-ratio $\ln \frac{p(x = x_i \mid y = c)}{p(x = x_i \mid y = c')}$ measures the relative prominence of token $x_i$ in class $c$ compared to class $c'$. This term effectively captures the divergence in token distributions between classes at the specific token $x_i$. Intuitively, tokens that frequently appear in class $c$ but are rare in other classes are considered more discriminative, and are therefore assigned higher $IS$ values. These tokens are likely to play a critical role in classification and should be perturbed less to preserve model utility when incorporating privacy mechanisms.

\para{Token Importance for Generation.} In generation tasks, where explicit supervision (\eg, class labels) is unavailable, we propose a label-free and computationally efficient metric, \ie, entropy-weighted attention aggregation, to estimate token importance. This method is grounded in the hypothesis that important tokens are not only highly attended to across the attention layers but are also attended to with high certainty, as reflected in the sharpness of the attention distribution. Formally, let $A^{(l,h)} \in \mathbb{R}^{N \times N}$ denote the attention matrix from the $h$-th head in the $l$-th layer, where $N$ is the sequence length. The entry $A^{(l,h)}_{ij}$ represents the attention weight from token $x_i$ (as the query) to token $x_j$ (as the key). We first compute the base importance score of token $x_i$ as the average attention it receives from all tokens: $S^{(l,h)}(x_i) = \frac{1}{N} \sum_{i=1}^{N} A_{ij}^{(l,h)}$. To quantify the attention certainty for each token $x_i$, we compute the entropy of its attention distribution over all keys:
\vspace{-0.15cm} 
\begin{equation}
	{E^{(l,h)}}({x_i}) =  - \sum\limits_{j = 1}^n {A_{ij}^{(l,h)}} \log (A_{ij}^{(l,h)}).    
\end{equation}
Since low entropy corresponds to high importance, we use the reciprocal of the entropy as a weight, \ie, $ W_{{\text{entropy}}}^{(l,h)}({x_i}) = \frac{1}{{{E^{(l,h)}}({x_i})}}$. We multiply the $S^{(l,h)}(x_i)$ by this weight to get the entropy-weighted single-head importance score:
\begin{equation} 
	S_{{\text{weighted}}}^{(l,h)}({x_i}) = {S^{(l,h)}}({x_i}) \cdot W_{{\text{entropy}}}^{(l,h)}({x_i}). 
\end{equation}

We average over all attention heads ($H$) and selected layers ($L_{\text{proc}}$) to obtain the raw token importance:
\begin{equation}
	IS_{\text{raw}}(x_i) = \frac{1}{|L_{\text{proc}}|} \sum_{l \in L_{\text{proc}}} \left( \frac{1}{H} \sum_{h=1}^{H} S_{\text{weighted}}^{(l,h)}(x_i) \right).
\end{equation}
To ensure consistency across different sequences, we normalize the raw scores using Z-score normalization: \begin{equation} 
	IS({x_i}) = \frac{{I{S_{raw}}({x_i}) - \mathbb{E}[I{S_{raw}}({x_i})]}}{{\sqrt {{\text{Var}}(I{S_{raw}}({x_i}))} }}.
\end{equation} 

\para{OPT-3 Guided Importance-Aware Noise Injection.} 
While the $d_\chi$-Privacy mechanism provides formal protection by injecting noise proportional to the sensitivity of the embedding function, its standard isotropic noise distribution is unaware of both task semantics and token-specific importance. To improve the utility of perturbed representations, we leverage the optimal perturbation $\bm{p}_i^*$ obtained from \textbf{OPT-3} and modulate the noise distribution using the token importance score $IS(x_i)$. Specifically, we shift the mean of the $d_\chi$ noise distribution to $\bm{p}_i$ to preserve the optimal perturbation direction. Meanwhile, we scale the noise magnitude based on the token-wise importance factor $S(x_i) = \frac{1}{1 + \exp(-IS(x_i))}$, such that more important tokens receive smaller perturbations. The final noise is sampled from a Laplace-like distribution:
\begin{equation}
	\bm{p}_i \sim \text{Lap}_d(\bm{p}) \propto \exp\left( -\frac{\epsilon}{S(x_i)\cdot \Delta\chi} | \bm{p} - \bm{p}_i^* |_2 \right),
\end{equation}
where $S(x_i) \in (0,1)$ adjusts the effective noise scale for token $x_i$, and $\bm{p}_i^*$ serves as the center of the distribution. \textit{This formulation preserves the core privacy guarantee of $d_\chi$-Privacy, since the noise distribution remains norm-based and properly calibrated.}

\section{Experimental Results}
\subsection{Experiment Settings}
To evaluate the performance of \texttt{PrivTune}, we conduct extensive experiments on five datasets. All experiments are developed using Python 3.9 and PyTorch 1.12 and evaluated on a server with an NVIDIA A100 Tensor Core GPU.

\para{Datasets.}
The effectiveness of the \texttt{PrivTune} is evaluated on both text classification and generation tasks. For the classification tasks, we use the Financial Phrasebank (FP) \cite{malo2014good} and Stanford Sentiment Treebank (SST) \cite{wang2019glue}, and the Blog Authorship Corpus~\cite{lyu2020differentially}. For the generation tasks, we use the question-answering datasets, \ie, TruthfulQA (TQA)~\cite{lin2022truthfulqa} and Fitness\_Unformatted (FU)~\cite{hazsylvia_fitness_unformatted_2024} datasets.

\begin{figure}[!t]
	\centering
	\subfigure{
		\includegraphics[width=0.47\textwidth]{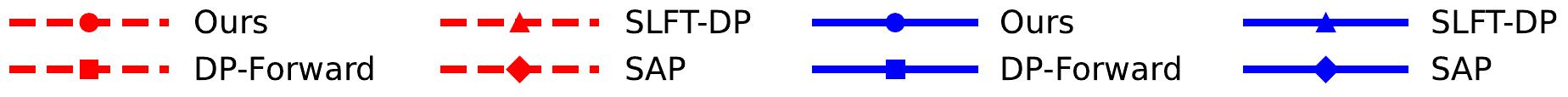}
	}
	\vspace{-19pt}
	\\
	\addtocounter{subfigure}{-1}
	\subfigure[RoBERTa, FP]{
		\includegraphics[width=0.3\linewidth]{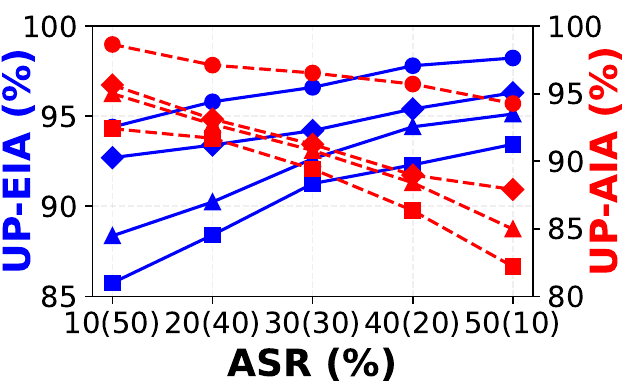}
	}\hfill
	\subfigure[RoBERTa, Blog]{
		\includegraphics[width=0.31\linewidth]{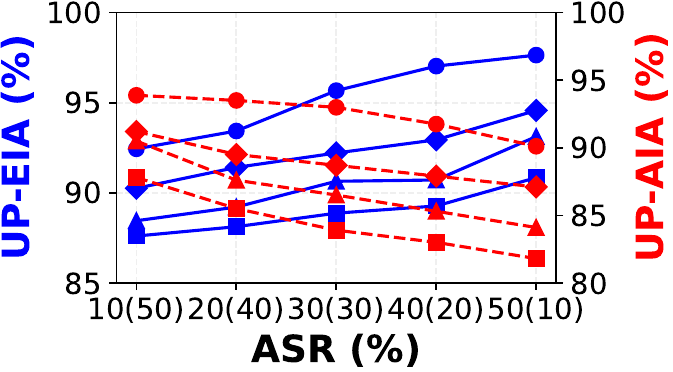}
	}\hfill
	\subfigure[RoBERTa, SST]{
		\includegraphics[width=0.3\linewidth]{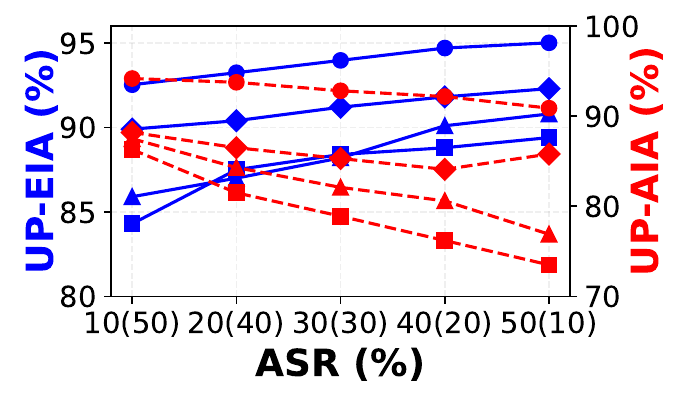}
	}
	\caption{Privacy-utility performance of RoBERTa. Note that the parts inside and outside ``()'' represent ASR under EIAs and AIAs respectively, and the same applies below.} \label{fig-2}
\end{figure}

\para{Models.}
RoBERTa-Large~\cite{liu2019roberta}, with 355 million parameters, Mistral-7B-v0.2~\cite{jiang2023clip}, with 7 billion parameters, and Llama-3~\cite{llama3modelcard}, with 8 billion parameters, are used as cloud models, both of which are publicly available at \href{https://huggingface.co/}{https://huggingface.co/}.

\para{Baselines.} \texttt{PrivTune} is evaluated against four baseline methods: 1) DP-Forward~\cite{du2023dp}, which directly perturbs the embedding matrix during the forward pass of the bottom model to meet strict DP requirements; 2) SLFT-DP~\cite{WANG2024Selective}, which applies sequence-level DP to perturb embeddings in the forward pass for privacy-preserving fine-tuning within the SLFT framework; 3) SAP~\cite{shen2025sap}, which integrates text privatization (\ie, $d_\chi$-Privacy) into an SL framework; 4) ScaleOT~\cite{yao2025scaleot}, which employs an offsite-tuning framework to achieve privacy preservation; and 5) FedBiOT~\cite{wu2024fedbiot}, which combines FL with offsite-tuning to enable secure and private fine-tuning.

\para{Privacy Inference Attacks.} To comprehensively evaluate the privacy performance of the \texttt{PrivTune}, we employ two types of state-of-the-art privacy inference attacks, \ie, EIAs and AIAs attacks~~\cite{song2020information,zhu2019deep,shokri2017membership,du2023dp,huang2024transferable}. Specifically, we follow their technical implementations (\ie, original settings) and utilize their source codes to reproduce three representative EIA (\ie, Attack-0--Attack-2) and three representative AIA (\ie, Attack-3--Attack-5) attacks for evaluating \texttt{PrivTune}. %Further details are provided in Section \S\ref{sec:related:prel} and Table \ref{tab:aia-eia-comparison}.

\para{Implementation Details.} We adopt the PEFT method, specifically LoRA~\cite{hulora}, in \texttt{PrivTune}, and employ the AdamW optimizer with a linear learning rate scheduler during fine-tuning, with an initial learning rate of $3 \times 10^{-4}$. By default, we set $l=3$, $n = 3$, $k = 2$, and $\delta = 0.6$ in our experiments. For evaluation, we use classification accuracy and ROUGE-L score to measure the Utility Performance (UP) on classification and generation tasks, respectively~\cite{li2023privacy}. Additionally, Attack Success Rate (ASR) is used to assess the defense capability.

\subsection{Numerical Results}

\para{Privacy-Utility Evaluation.}
Since \texttt{PrivTune} uses the $d_\chi$-Privacy mechanism different from DP techniques, it is difficult to directly compare it with baselines under different noise scales (\ie, $\epsilon$ in DP). For this reason, we compare the UP under different ASRs to illustrate the trade-off between privacy and utility. In addition, since ScaleOT and FedBiOT do not use DP-like techniques to protect privacy, we directly report the ASR. For clarity, we compute the average UP across three EIAs and three AIAs attacks on five datasets, using three LLMs.

\begin{figure*}[!t]
	\centering
	\subfigure{
		\includegraphics[width=1\linewidth]{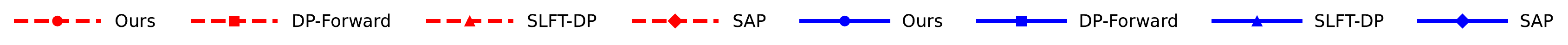}
	}
	\vspace{-20pt}
	\\
	\addtocounter{subfigure}{-1}
	\subfigure[Mistral-7B, FP]{
		\includegraphics[width=0.18\linewidth]{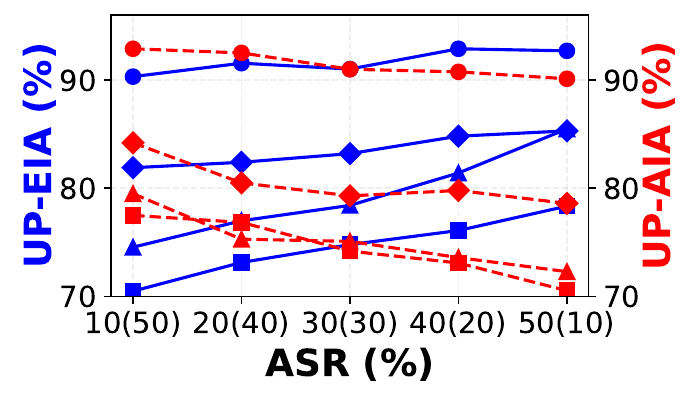}
	}
	\subfigure[Mistral-7B, Blog]{
		\includegraphics[width=0.18\linewidth]{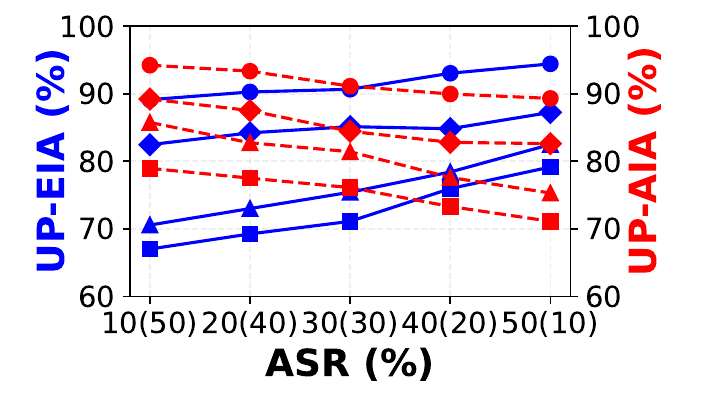}
	}
	\subfigure[Mistral-7B, SST]{
		\includegraphics[width=0.18\linewidth]{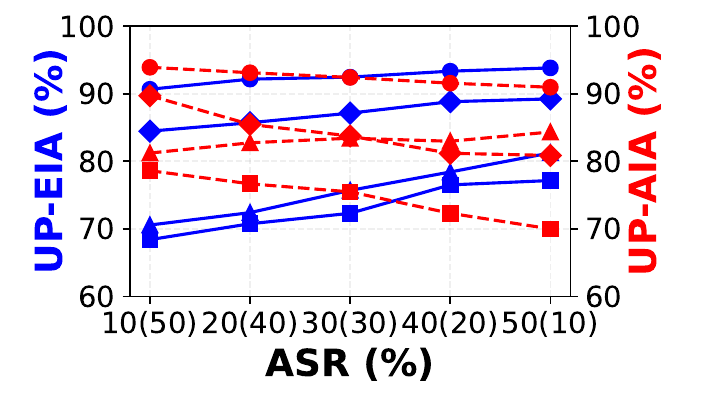}
	}
	\subfigure[Mistral-7B, TQA]{
		\includegraphics[width=0.18\linewidth]{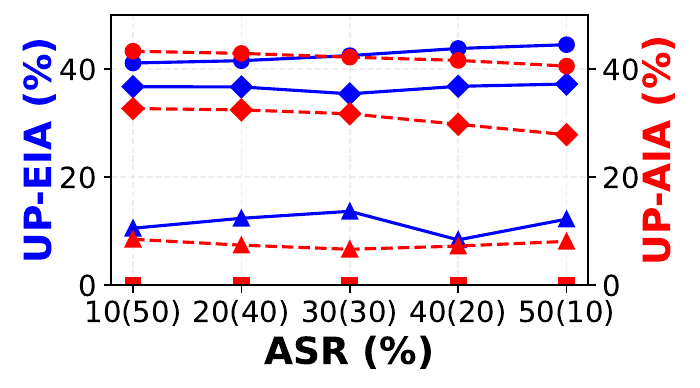}
	}
	\subfigure[Mistral-7B, FU]{
		\includegraphics[width=0.18\linewidth]{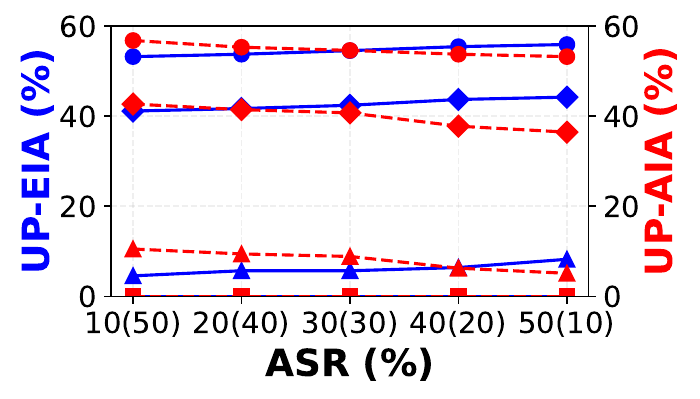}
	}
	\caption{Privacy-utility performance of different methods on Mistral-7B.} \label{fig-3}
\end{figure*}

\begin{figure*}[!t]
	\centering
	\subfigure{
		\includegraphics[width=\linewidth]{fig-line-legend-2.pdf}
	}
	\vspace{-20pt}
	\\
	\addtocounter{subfigure}{-1}
	\subfigure[Llama-3, FP]{
		\includegraphics[width=0.18\linewidth]{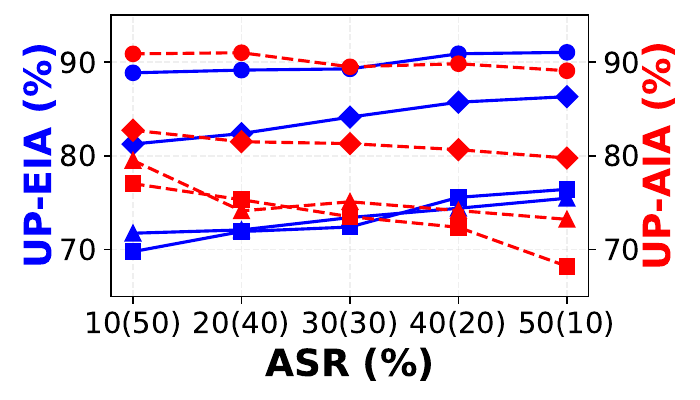}
	}\hfill
	\subfigure[Llama-3, Blog]{
		\includegraphics[width=0.18\linewidth]{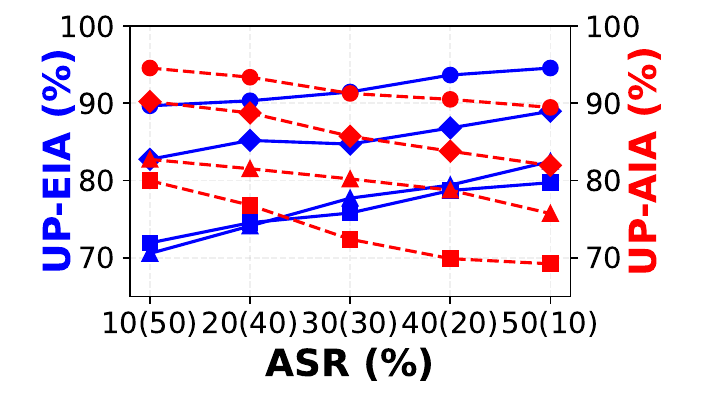}
	}\hfill
	\subfigure[Llama-3, SST]{
		\includegraphics[width=0.19\linewidth]{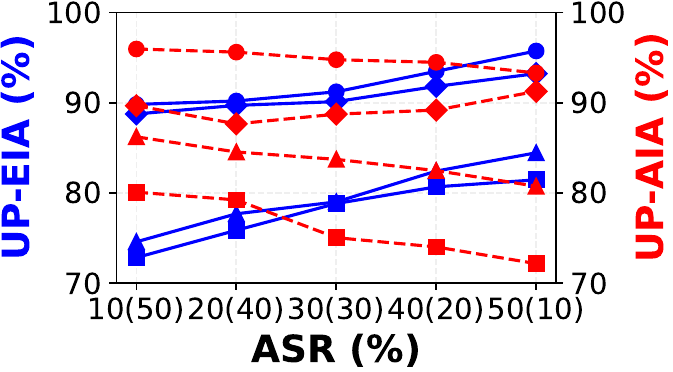}
	}\hfill
	\subfigure[Llama-3, TQA]{
		\includegraphics[width=0.19\linewidth]{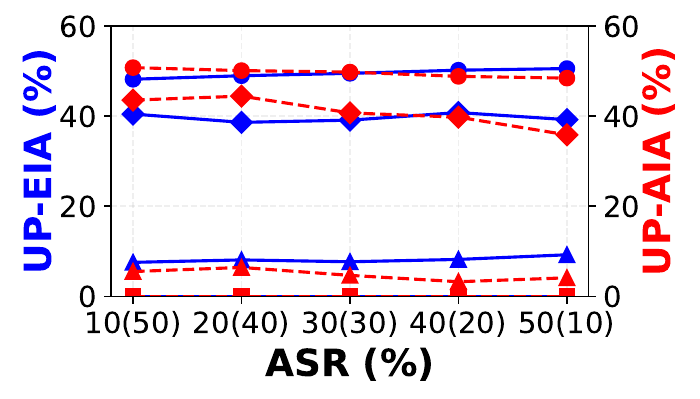}
	}\hfill
	\subfigure[Llama-3, FU]{
		\includegraphics[width=0.18\linewidth]{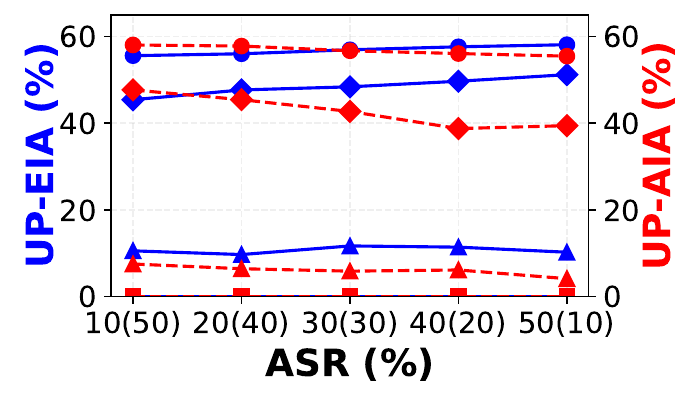}
	}
	\caption{Privacy-utility performance of different methods on Llama-3.} \label{fig-4}
\end{figure*}

\begin{figure}[!t]
	\centering
	\subfigure{
		\includegraphics[width=0.7\linewidth]{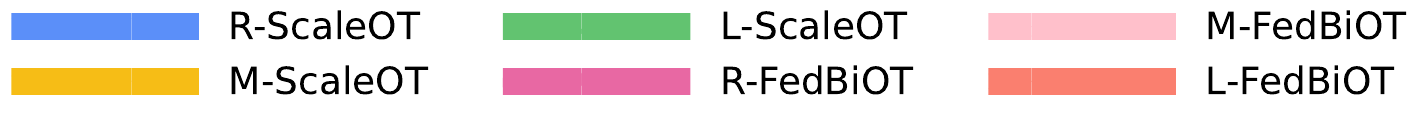}
	}
	\vspace{-10pt}
	\\
	\addtocounter{subfigure}{-1}
	\subfigure[EIAs]{
		\includegraphics[width=0.3\linewidth]{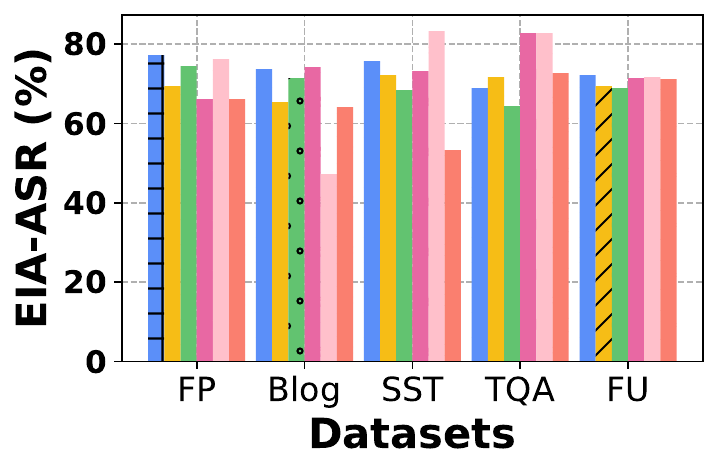}
	}\hfill
	\subfigure[AIAs]{
		\includegraphics[width=0.3\linewidth]{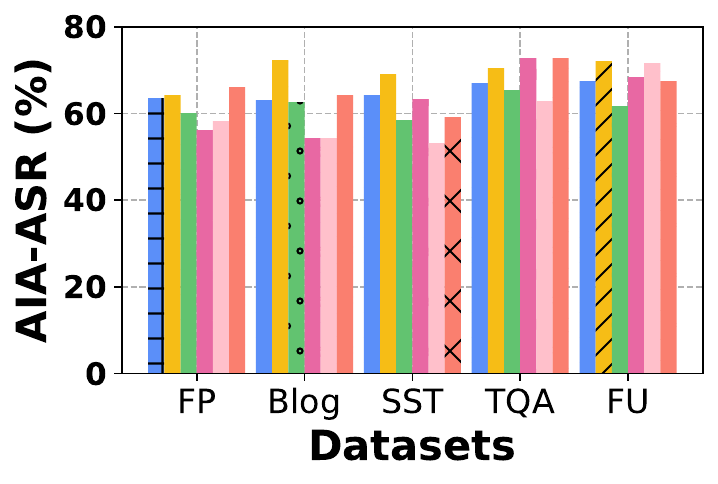}
	}\hfill
	\subfigure[UP]{
		\includegraphics[width=0.3\linewidth]{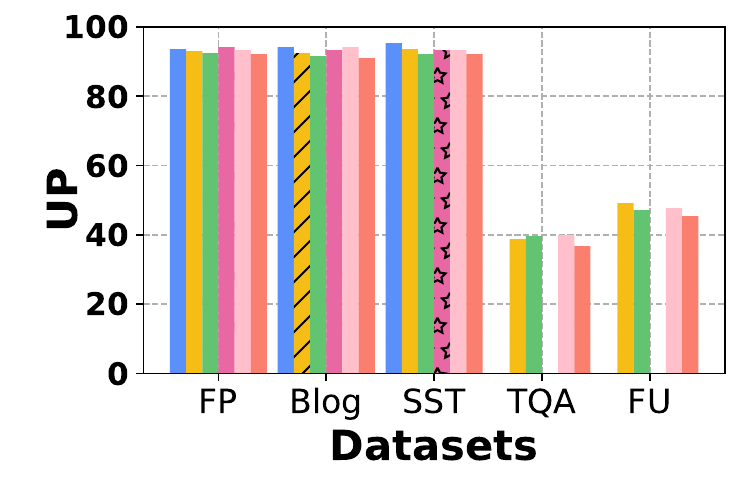}
	}
	\caption{Privacy-utility performance of ScaleOT and FedBiOT.} \label{fig-5}
\end{figure}

\textit{Privacy-Utility Evaluation on Classification Tasks.} We present the privacy-utility trade-off results across three LLMs and three classification datasets. Specifically, Figs. \ref{fig-2}, \ref{fig-3}, and \ref{fig-4} show the average UP of \texttt{PrivTune}, DP-Forward, and SAP using RoBERTa, Mistral-7B, and Llama-3, respectively, under varying ASR for both EIAs and AIAs. Fig. \ref{fig-5} reports the ASR and UP of ScaleOT and FedBiOT on the same models and datasets. We find that \texttt{PrivTune} achieves a superior privacy-utility trade-off compared to DP-based baselines, delivering higher UP at the same ASR level. For example, on the SST dataset with the RoBERTa model under EIAs, \texttt{PrivTune} achieves a UP of 92.53\% (drop 3.33\% in UP) at an ASR of 10\%, significantly outperforming other DP-based methods. On the FP dataset with Llama-3, \texttt{PrivTune} attains a comparable UP of 95.32\% relative to ScaleOT, while reducing ASR from approximately 50\% to just 10\% under AIAs. These results highlight \texttt{PrivTune}’s effectiveness in achieving strong privacy protection without sacrificing utility.

\textit{Privacy-Utility Evaluation on Generation Tasks.} We also evaluate the privacy-utility trade-off of different baselines on two generative datasets. Since RoBERTa does not support generation, experiments are conducted only with Mistral-7B and Llama-3. Figs. \ref{fig-3}, \ref{fig-4}, and \ref{fig-5} present the results for \texttt{PrivTune} and the baselines on these datasets. We find that while DP-based methods achieve strong privacy protection, they severely degrade generation performance, yielding ROUGE-L scores close to zero, largely due to sequence-level perturbations that disrupt autoregressive text generation. Meanwhile, ScaleOT and FedBiOT still suffer from high ASR values, indicating inadequate privacy protection. In contrast, \texttt{PrivTune} using the Llama-3 model not only supports effective text generation but also achieves a UP of 60 on the TQA dataset at an ASR of 10\% under EIAs, demonstrating the effectiveness of its token perturbations in preserving utility while enhancing privacy.

\para{Computational and Communication Cost Evaluation.} Table \ref{tab:example-1} demonstrates the low cost of \texttt{PrivTune} using Llama-3 on the SST dataset. Compared to ScaleOT and FedBiOT, its communication cost is only $1/2$ of theirs. Additionally, we find that the computational cost of \texttt{PrivTune} is comparable to that of DP-Forward, SLFT-DP, and SAP, with no additional computational burden, consistent with our theoretical analysis (see \S\ref{noise}). Overall, \texttt{PrivTune} is cost-effective and well-suited for deployment in edge device environments.

 \begin{table}[!t] 	\centering 	
 	\caption{Comparison of comp. and comm. cost per epoch.}\label{tab:example-1} 	
 	\vspace{-0.05in} 	
 	\resizebox{0.4\textwidth}{!}{% 		
 		\begin{tabular}{lcccc} 			
 			\toprule
 			\textbf{Method} & \textbf{\makecell{Memory\\(GB)}} & \textbf{\makecell{Comm. Cost\\(MB)}} & \textbf{\makecell{Comp. Cost\\(GFLOPs/token)}} & \textbf{\makecell{Running Time\\(s)}} \\
 			\midrule		
 			DP-Forward & 15.2 & 2040.02 & 6.34 & 867.24 \\ 			
 			SLFT-DP & 14.8 & 2040.02 & 5.41 & 656.34 \\ 			
 			SAP & 14.7 & 2040.02 & 5.21 & 668.18 \\ 			
 			ScaleOT & 31.5 & 4420.68 & 10.21 & 1415.62 \\ 			
 			FedBiOT & 33.2 & 4430.05 & 10.33 & 1118.91 \\ 			
 			\textbf{Ours} ($l=3$) & \textbf{6.8} & \textbf{2040.02} & \textbf{3.27} & \textbf{512.34} \\ 			
 			\textbf{Ours} ($l=4$) & \textbf{7.7} & \textbf{2872.05} & \textbf{4.19} & \textbf{567.38} \\ 			
 			\textbf{Ours} ($l=5$) & \textbf{8.6} & \textbf{3704.08} & \textbf{4.87} & \textbf{620.83} \\ 			
 			\bottomrule 	\end{tabular}} 	
 
 	\end{table}

\para{Parameter Sensitivity Analysis.} Next, we explore how the number of split layers, hops, and the $\epsilon$ affect the UP and ASR of \texttt{PrivTune} on the SST dataset using the Llama-3 model.

\textit{Impact of the Privacy Parameter $\epsilon$.} In Tables \ref{tab-3} and \ref{tab-4}, we explore the impact of different $\epsilon$ on the privacy-utility trade-off of \texttt{PrivTune} under EIAs, in terms of different $l$ and $n$. We observe that smaller values of $\epsilon$ result in lower UP but provide stronger privacy protection (lower ASR), while larger $\epsilon$ leads to weaker privacy but better utility. On the SST dataset, we find that $\epsilon = 10$ offers a better balance, achieving a UP of 89.81\% (drop 7.46\% in UP) with an ASR of only 8.28\%.

\textit{Impact of the Number of Split Layers.} Table \ref{tab-3} shows that increasing the $l$ in the bottom model enhances privacy, making it more difficult for attackers to infer the input text from the $\bm{h}$. Even without \texttt{PrivTune}, the ASR drops to 52.05\% when the bottom model includes 5 layers. We observe that as $l$ increases, the UP of \texttt{PrivTune} decreases slightly, while the ASR drops significantly. Therefore, service providers should aim to select a larger $l$ within the resource constraints of the user device to achieve a better privacy-utility trade-off.

\textit{Impact of the Number of Hops.} Table \ref{tab-4} presents the impact of the number of hops on the privacy-utility performance of \texttt{PrivTune} under EIAs. It shows that as the $n$ increases, the UP of \texttt{PrivTune} drops significantly, while the ASR only decreases slightly. For instance, compared to $n=3$, when $n=5$, the UP decreases by 5.7\%, whereas ASR only drops by 2.67\%. This indicates that a higher $n$ leads to greater loss of token semantics, thereby degrading performance. Hence, $n=3$ strikes a reasonable balance between privacy and utility.

\begin{table}[!t]
	\centering
	\caption{The impact of the number of split layers $l$.} 
	\label{tab-3}
	\vspace{-0.05in}
	\resizebox{0.4\textwidth}{!}{%
		\begin{tabular}{c|c|ccccccc} 
			\toprule 
			\multirow{2}{*}{\# of Split Layers}  & \multirow{2}{*}{Metric} &  \multicolumn{7}{c}{Privacy Parameter $\epsilon$}   \\
			\cline{3-9}  
			& & 10 & 20 & 30 & 40 & 60 & 80  & None      \\    
			\midrule    
			\multirow{2}{*}{3}  &  ASR & 8.28 & 21.03 & 29.46 & 41.18 & 45.79 & 51.42  & 82.64    \\  	
			% \cline{2-9} 
			& UP & 89.81 & 90.19 & 91.21 & 93.48 & 95.03 & 95.74 & 96.27 \\ 
			\midrule   
			
			\multirow{2}{*}{4} & ASR & 8.15 & 10.89 & 25.23 & 30.92 & 35.51 & 41.15 & 72.31 \\
			& UP & 87.05 & 89.92 & 91.08 & 92.20 & 93.12 & 94.25 & 95.25 \\
			\midrule
			
			\multirow{2}{*}{5} & ASR & 7.02 & 9.75 & 19.01 & 25.65 & 33.23 & 38.88 & 52.05 \\
			& UP & 85.32 & 85.75 & 87.51 & 90.02 & 92.93 & 93.07 & 94.76 \\
			\bottomrule 
		\end{tabular}
	}
\end{table}

\begin{table}[!t]
	\centering
	\caption{The impact of the number of hops $n$.} 
	\label{tab-4}
	\vspace{-0.05in}
	\resizebox{0.4\textwidth}{!}{%
		\begin{tabular}{c|c|ccccccc} 
			\toprule 
			\multirow{2}{*}{\# of Hops $n$}  & \multirow{2}{*}{Metric} &  \multicolumn{7}{c}{Privacy Parameter $\epsilon$}   \\
			\cline{3-9}  
			& & 10 & 20 & 30 & 40 & 60 & 80  & None      \\    
			\midrule    
			\multirow{2}{*}{3}  &  ASR & 8.28 & 21.03 & 29.46 & 41.18 & 45.79 & 51.42  & 82.64    \\   	
% \cline{2-9} 
			& UP & 89.81 & 90.19 & 91.21 & 93.48 & 95.03 & 95.74 & 96.27 \\  
            \midrule   
			\multirow{2}{*}{4}  &  ASR & 7.79 & 11.01 & 25.74 & 28.14 & 33.87 & 40.97 & 81.44 \\ 
			% \cline{2-9}
			& UP & 87.16 & 88.74 & 89.51 & 91.25& 92.89 &94.13 & 95.25 \\ 
			\midrule   
			\multirow{2}{*}{5}  &  ASR & 7.45 & 10.72 & 24.58 & 27.06 & 31.83 & 38.88 & 79.97       \\ 
			% \cline{2-9} 
			& UP & 82.45 & 84.01 & 85.46 & 86.51 & 88.77 & 89.06 & 90.78 \\ 
			\bottomrule 
		\end{tabular}
	}
\end{table}

\para{Ablation Analysis.} Finally, we examine the impact of key components, \ie, OPT-3, $d_\chi$-Privacy, and the $IS$, on the performance of \texttt{PrivTune}. Specifically, we use the RoBERTa model on the Blog dataset to evaluate UP and ASR under different component combinations. The results, shown in Table \ref{tab:ablation_study}, indicate that using OPT-3 or $d_\chi$-Privacy alone leads to similar UP (\ie, 88\% - 89\%), but $d_\chi$-Privacy alone achieves a lower ASR compared to OPT-3. This is because OPT-3 lacks formal privacy guarantees. Moreover, we find that incorporating $IS$ effectively enhances utility. Overall, the combined integration of all components achieves the best privacy-utility trade-off.
\begin{table}[htbp]
	\centering
	\small
	\caption{Ablation results.}
	\vspace{-0.05in}
	\label{tab:ablation_study}
	\resizebox{0.4\textwidth}{!}{
		\begin{tabular}{lccccc}
			\toprule
			\textbf{Model Variant} & \textbf{OPT-3}  &\textbf{$d_\chi$-Privacy}& \textbf{$IS$}&\textbf{UP}&\textbf{ASR} \\
			\midrule
			SLFT  & \ding{55} & \ding{55}& \ding{55} &98.29  & 79.84 \\
			
			+ \textbf{OPT-3}   & \checkmark& \ding{55}& \ding{55}& 89.14 & 12.55 \\
			
			+ $d_\chi$-Privacy& \ding{55}& \checkmark&\ding{55} & 88.07 & 11.84 \\
			
			+ $IS$-Privacy & \ding{55}&  \checkmark& \checkmark& 91.53 & 12.69\\
			
			+ $IS$-\textbf{OPT-3} & \checkmark&  \ding{55}& \checkmark& 91.66 & 13.54\\
			
			+ All & \checkmark&  \checkmark& \checkmark& \textbf{92.44} & \textbf{8.76}\\
			\bottomrule
	\end{tabular}}
\end{table}

\subsection{Discussion and Future Work}
\para{Expanding to Larger LLMs and Edge Use Cases.} \texttt{PrivTune}’s success with RoBERTa suggests promise that future work will adapt it to larger LLMs (\eg, 70B-parameter models) and edge settings, leveraging their architectures and data patterns to boost real-world utility.

\para{Dynamic Threat Adaptation.} Building on token-importance scaling, integrating threat detection to adjust noise dynamically will strengthen privacy for users while preserving utility, enhancing resilience to evolving attacks.

\section{Conclusion}\label{sec:conclusion}
We proposed \texttt{PrivTune}, a privacy-preserving fine-tuning framework for LMaaS, which integrates a token importance-aware $d_\chi$-Privacy mechanism and a defense-utility optimization strategy. By combining SL and PEFT, \texttt{PrivTune} enables resource-constrained users to access various LLM adaptation services while achieving superior privacy-utility trade-offs via its $d_\chi$-Privacy design. Experimental results demonstrate that, using the Llama-3 model on the SST dataset, \texttt{PrivTune} reduces the ASR to only 10\% with just a 3.33\% drop in UP.

\section*{ACKNOWLEDGMENT}
We thank all anonymous reviewers for their constructive comments. Cong Wang was supported in part by the Hong Kong Research Grants Council under Grants CityU 11218322, 11219524, R6021-20F, R1012-21, RFS21221S04, C2004-21G, C1029-22G, C6015-23G, and N\_CityU139/21 and in part by the Innovation and Technology Commission (ITC) under the Joint Mainland-Hong Kong Funding Scheme (MHKJFS) under Grant MHP/135/23. Cong Wang was also supported by the InnoHK initiative, the Government of the HKSAR, and the Laboratory for AI-Powered Financial Technologies (AIFT). Chengjun Cai was supported in part by the National Nature Science Foundation of China under Grant 6220239 and by the Guangdong Basic and Applied Basic Research Foundation under Grant 2023A151514013.

\bibliography{refs}
\bibliographystyle{ieeetr}
\end{document}